\documentclass[dvips]{acta}
\usepackage{lscape,epsfig}
\usepackage{amssymb}
\usepackage{amsmath}
\DeclareSymbolFont{ppa}{OT1}{ppl}{m}{it}
\DeclareMathSymbol{\vv}{\mathalpha}{ppa}{'166}

\newfont{\hb}{rphvb at 10pt}%bezszeryfowe p??grube
\newfont{\hbo}{rphvbo at 10pt}%bezszeryfowe p??grube kursywa
\newfont{\bitt}{rptmbi at 12pt}%p??gruba kursywa (tytu? artyku?u)
\newfont{\bits}{rptmbi at 11pt}%p??gruba kursywa (tytu?y rozdzia??w)

\SetPages{0}{0}

\SetVol{63}{2013}

\begin{document}
%Zwarte naglowki, jeden wiersz, appendix
\newcommand{\TabApp}[2]{\begin{center}\parbox[t]{#1}{\centerline{
  {\bf Appendix}}
  \vskip2mm
  \centerline{\small {\spaceskip 2pt plus 1pt minus 1pt T a b l e}
  \refstepcounter{table}\thetable}
  \vskip2mm
  \centerline{\footnotesize #2}}
  \vskip3mm
\end{center}}

%Zwarte naglowki, jeden wiersz
\newcommand{\TabCapp}[2]{\begin{center}\parbox[t]{#1}{\centerline{
  \small {\spaceskip 2pt plus 1pt minus 1pt T a b l e}
  \refstepcounter{table}\thetable}
  \vskip2mm
  \centerline{\footnotesize #2}}
  \vskip3mm
\end{center}}

%Zwarte naglowki, dwa wiersze
\newcommand{\TTabCap}[3]{\begin{center}\parbox[t]{#1}{\centerline{
  \small {\spaceskip 2pt plus 1pt minus 1pt T a b l e}
  \refstepcounter{table}\thetable}
  \vskip2mm
  \centerline{\footnotesize #2}
  \centerline{\footnotesize #3}}
  \vskip1mm
\end{center}}

\newcommand{\MakeTableApp}[4]{\begin{table}[p]\TabApp{#2}{#3}
  \begin{center} \TableFont \begin{tabular}{#1} #4
  \end{tabular}\end{center}\end{table}}

%Zwarte naglowki, jeden wiersz
\newcommand{\MakeTableSepp}[4]{\begin{table}[p]\TabCapp{#2}{#3}
  \begin{center} \TableFont \begin{tabular}{#1} #4
  \end{tabular}\end{center}\end{table}}

%Zwarte naglowki, jeden wiersz
\newcommand{\MakeTableee}[4]{\begin{table}[htb]\TabCapp{#2}{#3}
  \begin{center} \TableFont \begin{tabular}{#1} #4
  \end{tabular}\end{center}\end{table}}

%Zwarte naglowki, dwa wiersze
\newcommand{\MakeTablee}[5]{\begin{table}[htb]\TTabCap{#2}{#3}{#4}
  \begin{center} \TableFont \begin{tabular}{#1} #5
  \end{tabular}\end{center}\end{table}}

\begin{Titlepage}
\Title{The Clusters AgeS Experiment (CASE): \\
 Variable Stars in the Globular Cluster M4
%\footnote{Based on data obtained at the Las Campanas Observatory.}%
}
\Author{J.~~K~a~l~u~z~n~y$^1$,
      ~~I.~B.~~T~h~o~m~p~s~o~n$^2$,
      ~~M.~~R~o~z~y~c~z~k~a$^1$,
      ~~and~~W.~~K~r~z~e~m~i~n~s~k~i$^1$}
{ $^1$Nicolaus Copernicus Astronomical Center, ul. Bartycka 18, 00-716 Warsaw, Poland\\
     e-mail: (jka, mnr, wk)@camk.edu.pl\\
  $^2$Carnegie Observatories, 813 Santa Barbara Street, Pasadena, CA 91101-1292, USA\\
     e-mail: ian@obs.carnegiescience.edu}
\Received{March 1, 2013}
\end{Titlepage}

\vspace*{7pt}
\Abstract
{Based on over 3000 $BV$ images of M4 collected in years 1995 -- 2009
 we obtain light curves of 22 variables, 10 of which are newly detected
 objects. We identify four detached eclipsing binaries and
 eight contact binaries.
%The $rms$ of our $V$-band photometry varies from
% $\sim$5 mmag at the turnoff to $\sim$100 mmag at the faintest surveyed
% stars with $V$ = 22 mag.\\
 Accurate periods are found for all but two variables.
Nineteen variables are proper-motion members of the cluster,
and the remaining three are field stars. Five variables are optical
 counterparts of X-ray sources.
 For one of the variables unassociated with X-ray sources we report a
 flare lasting for about 90~min and reaching an amplitude $\Delta V=
 0.11$~mag.\\
 One of the new contact binaries has a record-low mass ratio
 $q=0.06$. Another four such systems show season-to season luminosity
 variations probably related to magnetic activity cycles, whose lengths
 are surprisingly similar to that of the solar cycle despite a huge
 difference in rotational periods. The location of contact binaries
 on the color-magnitude diagram of M4 strongly suggests that at least 
 in globular clusters
 the principal factor enabling EW systems to form from close but detached
 binaries is stellar evolution.
 We identify 46 blue and yellow stragglers in M4 and discuss their properties.
 We also derive a map of the differential extinction in the
 central part of M4, and determine the reddening of a selected reference
 region, $E(B-V)=$0.392 mag.
}
{\it globular clusters: individual: M4
             -- blue stragglers -- binaries: eclipsing -- stars: variables}
\section {Introduction}
M4~=~NGC~6121 is located about 1.8~kpc away from the Sun (Kaluzny et al. 
2013 and references therein) in an uncrowded stellar field at a galactic
latitude of $+16$~deg. These parameters coupled with its low
concentration (core and half-light radii of 1.2 and 4.3
arcmin, respectively (Harris
1996, 2010 edition)) make it an easy and attractive target for
detailed studies, in particular, for a search for variable
stars. The online catalog of Clement et al. (2001; updated in 2009)
contains 80 variables found in the M4 field, of which 50 are either
unquestionable or likely RR~Lyr stars. The CASE group conducted an
extensive search for detached eclipsing binaries in M4, and a
detailed analysis of three such system has recently been published
by  Kaluzny et al. (2013). In two earlier contributions (Kaluzny,
Thompson \& Krzeminski 1997; Mochejska et al. 2002) we reported the
detection of 15 variables which  
were located in the blue-straggler region or near the main sequence
(MS) on the cluster color-magnitude diagram (CMD). 
The present paper is a continumation of those 
studies.

We present the results of the
analysis of over 3000 $BV$ images collected at Las Campanas
Observatory in the years 1995 -- 2009.
Section 2 contains a brief report on the observations and explains the
methods used to calibrate the photometry. Section 3 is devoted to a
measurement of the differential extinction across the M4 field along
with the reddening $E(B-V)$ of a selected reference region. The new
variables are described in Section 4, and Section 5 is devoted to
a discussion of the population of blue/yellow straggler stars in M4.
A summary of the paper is containined in Section 6.
\section {Observations and Photometric Reductions}

The cluster was observed on the 2.5-m du Pont telescope equipped with  the TEK5
2K$^2$ CCD camera. All images were taken with the same set of filters at a
scale of 0.259~$^{\prime\prime}$ per pixel.
Most of the frames cover an $8.^{\prime}84\times 8.^{\prime}84$ field
centered 44$^{\prime\prime}$ south and 147$^{\prime\prime}$
east of the cluster center. Observations
aimed at covering eclipses of some binary systems involved a subfield of
$8.^{\prime}84\times 4.^{\prime}42$ centered 31~$^{\prime\prime}$
south and 36~$^{\prime\prime}$ east of the cluster center. These subrastered
images constitute about half of the observations. A detailed log of the observations
is accessible at the CASE web page \footnote{http://case.camk.edu.pl}.

The data were collected on 51 nights between UT~June 1, 1995 and UT~June 30, 2009.
The images were taken in $V$ and $B$ filters at a median seeing of 0.99 and 1.07
arcsec, respectively, with average exposures of 47 s for $V$ and 102 s for $B$.

In total, we obtained 2447 useful frames in $V$ and 672 in $B$. The
light curves were extracted with a modified version of the image
subtraction utility ISIS V2.1 (Alard \& Lupton 1998; Alard 2000).
The step involving interpolation of images was performed using the
IRAF\footnote{IRAF is distributed by the National Optical Astronomy
Observatory, which is operated by the Association of Universities
for Research in Astronomy, Inc., under a cooperative agreement with
the National Science Foundation.} tasks $immatch.geomap$ and
$immatch.geotran$ instead of the $interp$ routine from the ISIS
package. Daophot, Allstar and Daogrow code (Stetson 1987, 1990)
was used to extract the profile photometry of point sources and to
derive aperture corrections for reference images. The reference
image in $V$ was constructed by combining eight individual frames
taken one after another, each exposed for 19~s. A total of seven  $B$ frames
were combined, each exposed for 60s. The seeing for the reference images
in $V$ and $B$ was 0.66 and 0.78 arcsec, respectively. For the
analysis, each frame was divided into  $4\times 4$ or $4\times 2$
segments to reduce the effects of PSF variability. 

The light
curves derived with ISIS were converted from differential counts to
magnitudes based on profile photometry and aperture corrections
determined separately for each segment of the reference images.
Instrumental magnitudes were transformed to the standard $BV$ system
as described in Section 2.1. The $V$-band light curves were extracted for
15240 sources. A plot of the $rms$ of the individual
measurements versus average magnitude for all of the
sources is shown in Fig.~1. The photometric accuracy is
about 4 mmag at $V=16.0$~mag, decreasing to 49~mmag at $V=20.0$~mag, 
and to $\sim$100 mmag at $V=22.0$~mag. We note that
depending on exposure time and seeing, some images of stars with
14.7$<V<16.0$ mag were saturated. The light curves of
these objects were carefully
filtered to remove points affected by saturation. All stars with
$V<14.7$ mag were overexposed on the reference $V$ image, making the
study of their light curves impossible.

\subsection{Calibration}
Accurate calibration of the photometry was essential for the analysis of three
detached eclipsing binaries described by Kaluzny et al. (2013). To achieve
the required accuracy, selected Landolt standards (Landolt 1992) were observed 
on several nights in different seasons. Two of those observations
%, 2 May 2002 and 5 May 2003,
proved to be of particularly high quality thanks to good seeing and stable
transparency. On UT~May 2, 2002 we observed 22 standards from five Landolt
fields (some fields were observed several times at different airmasses). In
total, 82 $B\&V$ measurements at air masses $1.07<X<2.00$ were collected.
On UT May 5, 2003 we observed 39 standards from seven fields. Again, some
fields were observed more than once, totalling  84 $B\&V$ measurements
at air masses $1.06<X<1.94$.
Aperture photometry was extracted with Daophot, and total magnitudes were
derived using Daogrow (Stetson 1987, 1990). Standard magnitudes of stars
from Landolt fields were taken from the catalog of Stetson (2000, January 2011
online edition). The following linear transformation from the instrumental
(small symbols) to the standard system was derived for UT May 5, 2003:
\begin{eqnarray}
{\rm v=V -0.0112(24)\times (B-V) + 0.1129(40)\times X + const,} \\
{\rm b=B -0.0580(27)\times (B-V) + 0.2054(39)\times X + const,} \\
{\rm b-v=0.9520(22)\times (B-V) + 0.0979(33)\times X + const.}
\end{eqnarray}
Fig. 2 shows residual differences between standard and recovered magnitudes
and colors, with $rms$ values of 0.0091, 0.0086 and 0.0071 for $V$, $B$ and $B-V$,
respectively. The residuals do not show any systematic
dependence on the color index. On the same night several frames of the M4 field
were collected at $X=1.0$ with sub-arcsecond seeing. These data were used to
obtain the calibrated $BV$ photometry of the cluster.

For the night of 2 May 2002, we obtained
\begin{eqnarray}
%{\rm v=V -0.0133\times (B-V) + 0.1225(X-1.25) + 0.7567,}
%             25                  30               15
%{\rm b=B -0.0658\times(B-V) + 0.2080(X-1.25) + 1.1132,}
%             27                  32               17
%{\rm b-v=0.9436\times(B-V) + 0.0904(X-1.25) + 0.0904,}
%            31                  37               20
{\rm v=V -0.0133(25)\times (B-V) + 0.1225(30)\times X + const,}\\
{\rm b=B -0.0658(27)\times (B-V) + 0.2080(32)\times X + const,}\\
{\rm b-v=0.9436(31)\times (B-V) + 0.0904(37)\times X + const.}
\end{eqnarray}
The $rms$ values of the residuals are 0.0077, 0.0077 and 0.0090,
for $V$, $B$ and $B-V$, respectively.
As in the previous case, the residuals do
not show any correlation with color. On the same night M4 was
observed at $X=1.25$ with sub-arcsecond seeing. Calibrated
$BV$ photometry was derived for cluster stars and compared to that
from May 5th, 2003. We find that color terms of the two
transformations agree with each other to within about one sigma.
The average differences of magnitudes and colors calculated for
$16<V<19$ mag are $\delta V = -0.0132$ mag and $\delta
(B-V)=-0.0104$ mag, with 2002 magnitudes brighter and colors redder.
The zero points of the finally adopted $BV$ photometry of reference
images were chosen as the averages of zero points defined by 2002 and
2003 photometry. The whole observed field
contained 83 secondary standard stars from the catalog of Stetson (2000).
The average residuals for these stars are $\Delta V=+0.016 \pm 0.0091$ mag
and $\Delta (B-V)=0.0235 \pm 0.0071$ mag, with our magnitudes brighter and
our colors bluer.

\section{Extinction and Color-Magnitude Diagram}

The interstellar extinction exhibits a significant variability
across the M4 field (Cudworth \& Rees 1990). Maps of differential
reddening $\delta E(B-V)$ were derived by Mochejska et al. (2002),
and more recently by Hendricks et al. (2012) and Monetti et al.
(2013). Hendricks et al.~also
show that the reddening law for the line of sight toward M4 has
the form $A_{V}/E(B-V)=3.76$, and that the average reddening in
a circle with a 10 arcmin radius around the cluster center is
$E(B-V)=0.37\pm 0.01$.

The quality of our photometry is high enough to attempt at an
independent determination of differential reddening in the observed
field. We follow the approach developed by Kaluzny \& Krzeminski
(1993) in an analogous study of the globular cluster NGC~4372, which
later on was reinvented by several other authors. The basic
assumption is that the intrinsic width of the principal CMD features
(main sequence, subgiant, giant and horizontal branch) is small, and
that the cluster is rich enough for these sequences to be densely
populated. One may then divide the observed field into segments and
derive the differential reddening by shifting their CMDs along the
reddening vector until they match each other. Needless to say, such
an approach works best when the cluster hosts just one population of
stars.

In the case of M4 we split the observed field into an $8\times 8$ mosaic of equal
circular subfields with a radius of 65 arcsec, each which were partially overlapping.
The differential reddening was derived in reference to a $150\times 150$~arcsec$^{2}$
square template region centered at $(\alpha, \delta)_{2000}$ = (245.89109,
-26.49646)~deg. The reddening in this area is nearly uniform, as indicated
by the narrowness of main-sequence and subgiant branch on the CMD. We cleaned the
template of field stars using the proper motion catalogue of Zloczewski et al.
(2012). From among the remaining cluster members, only those populating main
sequence and subgiant branch in a magnitude range 15 mag $<V<$ 20 mag were used for
the analysis.
The merit function used to derive $\delta E(B-V)$ was defined as follows: i) Select
a subfield. ii) For each template star select from that subfield stars located
closer than 0.04 mag on the ($B-V$, $V$) plane. iii) Derive the average distance
of those stars from the template star. iv) Sum squared average distances over
all stars from the template region. v) Minimize the result as a function of
$\delta E(B-V)$.

Fig. 3 is a schematic map of the differential extinction derived for
the above quoted reddening law of Hendricks et al. (2012). The map
in a tabular form is available in the electronic version of this
paper. The derived values of $\delta E(B-V)$ range from -0.053 to
+0.011, with negative values corresponding to higher reddening. To
correct the CMD of M4 for the effects of differential reddening we
interpolated $\delta E(B-V)$ for each star. The results of this
procedure are illustrated in Fig. 4, whose left panel includes all
stars measured on $B$ and $V$ reference images. A sample of stars
with $V<14$ mag and photometry based on a few short-exposure
frames taken at a seeing of about $\sim$1~arcsec is shown to
indicate the location of the red giant and horizontal branches. The
nearly vertical feature at $(B-V)\approx 1$ mag between
$V=22$ mag and $V=19$ mag is composed of stars belonging to Milky
Way's bulge. The middle panel shows the same CMD corrected for
differential reddening and cleaned of objects with relatively poor
photometry, which we identified based on the analysis of
$V/\sigma_{V}$ and $V/\sigma_{B-V}$ relations. In the right panel only
proper motion members of M4 are shown, selected from the
middle-panel sample. Note that the proper-motion membership survey
of Zloczewski et al (2012) is limited to stars brighter than 21 mag
in $V$, and it is very incomplete for bright stars with $V<13.5$
mag.

The reddening for the reference field was derived
%in a different way,
using the
low-extinction globular cluster NGC~6362 as an intermediary. For the latter we
obtained $BV$ photometry using the same equipment and the same methods as described
above. For both NGC~6362 and the reference field of M4 we measured average $B-V$
colors at the turnoff, obtaining $0.5404\pm 0.0004$ ($rms=0.0113$, 665 stars)
and $0.843\pm 0.001$ ($rms=0.0099$, 85 stars), respectively. The difference in
turnoff color, equal to $0.303\pm 0.001$, can result from differences in reddening,
metallicity and age between the two clusters. According to Dotter et al. (2010), M4
and NGC~6362 are coeval at about 12.5~Gyr. This finding is confirmed by our data, since
the CMDs of the two clusters match very well when offsets $\Delta(B-V)=0.303$ mag and
$\Delta V=1.97$ mag are applied.
%If the clusters had the same metallicity, they would have the same intrinsic turnoff
%color.
\begin{table}
\begin{center}
\caption{Equatorial coordinates of M4 variables identified within the present 
survey}
\begin{tabular}{|l|c|c|l|c|c|}
\hline
ID &  RA(J2000)& Dec(J2000) & ID &  RA(J2000)& Dec(J2000) \\
      &  [deg]    & [deg]   &    &  [deg]    & [deg]      \\
\hline
44 &  245.83760 & -26.55700 & 56 &  245.89282 & -26.49887 \\
46 &  245.94653 & -26.53236 & 58 &  245.89228 & -26.52629 \\
47 &  245.85642 & -26.48659 & 59 &  245.91317 & -26.49859 \\
48 &  245.90328 & -26.52891 & 60 &  245.94794 & -26.53663 \\
49 &  245.89301 & -26.53384 & 61 &  245.92637 & -26.55500 \\
50 &  245.88047 & -26.53015 & 62 &  245.88582 & -26.60363 \\
51 &  245.88830 & -26.51918 & 63 &  245.86390 & -26.54075 \\
52 &  245.88111 & -26.51607 & 64 &  245.83658 & -26.51727 \\
53 &  245.91034 & -26.53635 & 65 &  245.86826 & -26.50608 \\
54 &  245.96223 & -26.57836 & 66 &  245.88431 & -26.52813 \\
55 &  245.94067 & -26.52127 & 68 &  245.91071 & -26.50860  \\
\hline
\end{tabular}
\end{center}
\vspace*{-0.3cm}
{\footnotesize Objects V57 and V67 are not
listed here as the first one turned out to be constant, and the second one  
is located outside the field analyzed in this paper. }

\end{table}

Three independent surveys (Zinn and West 1984; Carretta et al. 2009;
Dotter et al. 2010) consistently indicate that the difference in
[Fe/H] between M4 and NGC~6362 amounts to 0.10, with the latter
being more metal rich. Based on Dartmouth Isochrones (Dotter et al.
2008) we find that for an age of 12.5~Gyr, and ${\rm [Fe/H]}=-1.1$, a
metallicity decrease $\Delta{\rm[Fe/H]} = 0.1$ causes the turnoff
color to decrease by 0.019 mag. According to Harris (1996, 2010
edition) NGC~6362 is reddened by 0.09 mag. This value seems to be a
weighted average of $E(B-V)=0.08$ mag from Reed, Hesser \& Shawl
(1998) and $E(B-V)=0.11\pm 0.03$ mag from Zinn (1985), with both these
estimates based on the integrated cluster light. The value we
adopt results from the SDF map of the foreground reddening published
by Schlegel et al. (1998), which at NGC~6362 coordinates gives
$E(B-V)=0.075$ mag. According to Schlafly et al. (2010), in regions
of low reddening the SDF values of $E(B-V)$ are overestimated by
14\%. As a result the reddening of NGC~6362 is reduced to 0.070 mag, 
and the reddening of our template field in M4 can finally be calculated 
as $E(B-V)=0.070+0.303+0.019=0.392$.

\section{Variables}
We derived $V$-band light curves for 15240 stars and examined them in
a search for variables with AoV and AOVTRANS algorithms (Schwarzenberg-Czerny
1996, Schwarzenberg-Czerny \& Beaulieu 2006) implemented in the TATRY
code. Ten new variables were identified. Our field also contains 13 
objects with $V>15.7$~mag which were previously identified as variables V44 
-- V57. Of these V57 turned out to be constant.
Following the numeration from earlier CASE papers, we named the new variables
V58-V56 and V68. Variable V67, discussed by Kaluzny et al. (2013), is not 
included here because it is located outside the observed field.  
Equatorial coordinates of all 22 confirmed variables are listed in Table 1.

\begin{table}
\begin{center}
\caption{Basic data of M4 variables identified within the present 
survey}
\begin{tabular}{|l|l|c|c|c|c|c|c|l|}
\hline
ID & P & $V_{max}$ & $B-V$ & $\Delta V$ &$\Delta E(B-V)$ &PM$^a$&Type$^b$\\
   &[d]& mag       &  mag  &  mag       & mag            &      &        \\
\hline
44 & 0.263584542(2) & 17.746 & 0.987& 0.19 & -0.020& Y & EW\\
46 & 0.0871525535(3)& 18.553 & 0.287& 0.05 & -0.005& Y & sdB\\
47 & 0.269874785(2) & 16.826 & 0.815& 0.27 & -0.005& Y & EW\\
48 & 0.28269398(2)  & 16.805 & 0.812& 0.31 & -0.005& U & EW CX15\\
49 & 0.297443902(1) & 17.087 & 1.290& 0.99 & -0.007& N & EW CX13\\
50 & 0.26599834(1)  & 17.251 & 0.830& 0.48 & -0.006& Y & EW\\
51 & 0.303682374(3) & 17.103 & 0.870& 0.45 & -0.001& Y & EW\\
52 & 0.7785087(15)  & 16.863 & 0.950& 0.13 & -0.003& Y & -\\
53 & 0.308448705(1) & 15.757 & 0.594& 0.23 & -0.007& Y & EW\\
54 & 0.25244661(1)  & 17.802 & 0.951& 0.24 & -0.008& Y & EB\\
55 & 0.310703451(1) & 16.722 & 0.809& 0.41 &  0.001& Y & EW\\
56 & -              & 14.646 & 1.154& 0.14 &  0.000& Y & CX5 CX9\\
%57 & -              & 17.545 & 0.834& 0.0  &  0.000& Y & const\\
58 & 0.2628216(1)   & 20.38  & -    & 0.35 & -0.001& U & CX1\\
59 & 0.71155962(4)  & 20.296 & 1.565& 0.41 &  0.000& Y & EA\\
60 & 0.370342607(7) & 19.297 & 1.206& 0.51 & -0.005& Y & EA\\
61 & 0.0413287(1)   & 15.683 & 0.590& 0.01 & -0.014& Y & SX\\
62 & 0.04062724(10) & 19.074 & 0.608& 0.07 & -0.040& N & SX\\
63 & -              & 17.670 & 0.101& 0.03 & -0.012& Y & sdB?\\
64 & 0.040953844(9) & 18.439 & 0.609& 0.05 & -0.008& N & SX\\
65 & 2.29304564(26) & 17.028 & 0.903& 0.40 & -0.006& U & EA CX30\\
66 & 8.11130346(85) & 16.843 & 0.878& 0.70 & -0.003& Y & EA\\
68 & 0.0380887(1)   & 15.238 & 0.615& 0.01 &  0.000& Y & SX\\
\hline
\end{tabular}
\end{center}
\vspace*{-0.3cm}
{\footnotesize Notes: $^a$ Membership status --    Y-member, N -field
star, U-unknown;
$^b$ EW - contact binary, EB - close eclipsing binary,
EA - detached eclipsing binary, SX - SX~Phe type pulsator, CX -
X-ray source from Bassa et al. (2004), sdB - hot subdwarf.\\
Variables V58-V66 and V68 are newly detected. Objects V57 and V67 are not 
listed here as the first one turned out to be constant, and the second one 
is located outside the field analyzed in this paper. }
\end{table}
The astrometric solution for the reference image in $V$ was found
based on positions of 273 UCAC3 stars (Zacharias et al. 2010). The
average residuals in RA and DEC between catalogued and recovered
coordinates amount to $0.00\pm 0.13$ and $0.00\pm 0.13$ arcsec,
respectively.
Finding charts for V44-V56 can be found in Kaluzny, Thompson
\& Krzeminski (1997). The finding chart for V63 was published by  Mochejska
et al. (2002), who, however, did not detect its variability, and classified
it as a blue object, named B3. Based on the photometric data described here and an
extensive spectroscopic survey, variables V65 and V66 were examined in detail
by Kaluzny et al. (2013), who also provide the relevant finding charts.
The remaining seven charts for the new variables are presented in~Fig.~5.

To check on the effect of blending on the photometry of our variables, we
examined HST/ACS images of M4 and the catalog of Anderson et al. (2008). V48
has three close visual companions located at distances of 0.21 to 0.84 arcsec.
We measured
profile photometry for this quadruplet in the fixed-position mode using positional information
extracted from the data of Anderson et al. (2008). The detached binary V65 has two
close visual companions at distances of 0.37 and 0.57 arcsec. Their presence was
deduced from the ground-based data, but accurate coordinates could only be determined
based on HST/ACS photometry. The case of V58 was discussed in detail by Kaluzny et al.
(2012). This $V\approx 20.5$ optical counterpart to
the X-ray source M4-CX1 (Bassa et
al. 2004) has a bright companion at a distance of 0.15 arcsec which
we originally
regarded to be the source of variability. A closer analysis unambiguously showed
that the fainter component of the blend is responsible for the observed variations.
As far as can be established based on HST/ACS data, the images of the remaining
variables are free from significant blending problems.

The basic properties of the variables are listed in Table 2. The
periods in Column 2 were derived with the help of the above
mentioned code TATRY. In the case of the eclipsing binaries the ($B-V$)
values in Column 3 correspond to the phase of maximum brightness.
For the remaining stars we provide median ($B-V$) from all available
measurements. Column 6 contains differential corrections to the
reference reddening of 0.392 mag. They are interpolated from the map
shown in Fig. 3. The membership status in Column 7 is taken from the
proper motion study of Zloczewski et al. (2012). It is worth noting
that in their Fig. 1 members of M4 are very clearly separated from
field stars. The classification of the variables as well as associated X-ray sources
discovered by Bassa et al. (2004) are given in Column 8. The optical
variables were identified with X-ray sources based on positional
coincidences. In all cases but one the identification was
unambiguous. The exception was V56, located between X-ray
sources CX5 and CX9 which are separated by $1.^{\prime\prime}3$.
Positions of variables on the CMD of M4 is shown in Fig. 6.

\subsection{Eclipsing binaries}
The sample of variables includes thirteen eclipsing binaries. All
but one of these are likely proper motion members of the cluster.
Their phased $V$-light curves for the variables are presented in 
Fig. 7. In four cases
the light curve indicates a  detached configuration. Variables V65
and V66 are located at the cluster turnoff. A 
detailed analysis of these two systems,
including a determination of absolute parameters and ages, was
presented in Kaluzny et al. (2013). The faint binary V59 is located 
$\sim$0.7 mag above MS which indicates that its mass ratio is close 
to unity.
The light curve of V60 shows noticeable curvature between the
eclipses. Despite the rather short orbital period of 0.37~d this
system is undoubtedly detached. This conclusion is based on a
preliminary analysis of the light curve performed with the PHOEBE
utility (Pr\^sa \& Zwitter 2005).
%analysis gives relative radii of components $r_{p}=31$ and $r_{s}=0.19$.

The remaining nine eclipsing variables are classified as contact binaries (EW).
All but V49 have proper motions consistent with cluster membership. Two systems,
V50 and V54, show a particularly large difference between the depths of primary
and secondary eclipses. Such light curves are called EB-type, and sometimes they
are observed in semidetached binaries. Since the eclipses of V50 and V54 are
partial, and the mass ratios are unknown, we cannot exclude that these two systems
are semidetached rather than contact. However, their short orbital periods favor
the second possibility (in particular, the light curves of V50 and V54 can be 
reproduced assuming a contact configuration and modest mass ratios of $\sim$0.3).

All but possibly one of the EW systems show season-to-season
variations of the average luminosity. Particularly large and
systematic variations occur at all orbital phases in V44, V48, V50
and V55, where the whole light curve varies by more than
0.1 mag (see Fig. 8). A similar behaviour is observed in V54 (see
Fig. 7), but the changes are less systematic. Similar effects  were
reported and discussed by Rucinski \& Paczynski (2002) in an EW
system observed during three consecutive seasons by the OGLE team,
and it would be certainly worthwhile to analyze in this respect the
sample of 569 contact binaries observed by OGLE during 14 seasons
(Kubiak, Udalski \& Szymanski 2006). Such  luminosity
variations can be explained by magnetic cycles analogous to the
Solar cycle. The length of the cycle is about 5 years in V44, 8-9
years in V48 and V55, and more than 9 years in V50. It is
interesting that while our EW systems rotate  two orders of
magnitude faster than the Sun, their magnetic activity periods do
not vastly differ from the solar period. Finally, we note that several
EW systems in M4 show evidence for orbital period variability, with
particularly rapid changes occurring in V47 and V50. A detailed
analysis of this effect is, however, beyond the scope of the present
paper.

Our survey of the contact binaries in M4 leads to two interesting 
conclusions. First, that EW systems
are absent among unevolved cluster stars. On the CMD of the cluster they
begin to appear about 1 mag below the turnoff, at $V\approx 17.8$ mag and $B-V\approx
0.95$ mag. These values correspond to $M_{\rm V}=5.0$ mag and $(B-V)_{0}=0.55$ mag,
which at an age of $\sim$12 Gyr implies primary masses of about 0.7~$M_{\odot}$.
Moreover, five out of the seven contact binaries belonging to the MS are located
within $\approx 0.3$~mag from the turnoff, and not a single such
system was detected among 12246 stars with $17.8<V<21$ mag whose
light curves we analyzed. Fig.~1 makes it obvious that we would
detect all EW variables down to $V\approx 22$~mag with light-curve
amplitudes larger than 50 mmag. And indeed there are numerous binaries
in M4: the right panel of Fig.~4 shows a well populated sequence
parallel to the MS which is most simply interpreted as a binary sequence.
These two findings indicate that at
least in globular clusters the principal factor enabling EW systems
to form from close but detached binaries is  stellar evolution.
A similar  conclusion results from the studies of variables in  M55 (Kaluzny
et al. 2010) and NGC~6752 (Kaluzny \& Thompson 2009). 
Apparently, even modest stellar evolution is more important in this respect than the
frequently invoked mechanism of magnetic breaking; see e.g. Stepien \& Gazeas
(2012) and references therein.

The second conclusion concerns the frequency of contact binaries
among evol\-ved MS stars in M4. W~UMa variables of spectral types F-G
are common in the nearby thin disc population. Rucinski (2006)
estimated that in the solar neighborhood their frequency relative to the 
MS is as high as 0.2\%. A similar frequency was found for the thick
disc population by Nef \& Rucinski (2008). In M4 we detected seven
contact binaries among 5652 upper MS stars with $16.5<V<18$ mag. The
corresponding relative frequency of occurrence, 0.12$\pm$0.05\%, is
consistent with that found for field stars.

The eclipses in V48 and V53 systems are total (Fig. 7). Mochnacki and Doughty (1972)
demonstrated that the totality of eclipses in a contact binary allows for a reliable
determination of two otherwise degenerate parameters: mass ratio $q$ and inclination
$i$. We analyzed the $V$-light curves of V48 and V53 with PHOEBE, obtaining $q=0.148$ and
$q=0.060$ for V48 and V53, respectively. The latter result is very interesting as it
sets a new record-low mass ratio, beating SX~Crv with $q=0.066\pm 0.003$ measured
spectroscopically by Rucinski et al. (2001). Mass ratios that low pose a challenge for
the current theory which predicts $q=0.09-0.08$ as the lower limit for tidally
stable contact binaries (Rasio 1995; Arbutina 2012). A detailed analysis of
V48 and V53  is deferred to a separate paper.

%
%To avoid complications related to long term changes of the
%average light level we selected for each star a small section of
%the whole available light curve. For V53 we used observations
%collected on a single  night of 2008 May 02. It
%provides continuous coverage for a span of 9h 50m and covers
%more than one orbital cycle.
% V53
%a q=0.0653 i=73.0d deg.
%b   0.06393 74.38171%
%%%      q      i
% V48 a  0.13639  81.29455
%     b  0.14796  83.18524
%     c  0.14906  84.29640
%

\subsection{Other variables}
Variable objects V53, V61 and V68 belong to the group of blue stragglers 
discussed in Section 4.3. The first one was described above. The remaining 
two are pulsating stars of SX~Phe type. They belong to the bluest of all 
M4 stragglers and are apparently located at the red boundary of the instability
strip for M4 metallicity of ${\rm [Fe/H]}=-1.16$ (Harris 1996, 2010 edition).
We note that the
analysis of blue straggler populations in several globular clusters
would allow for an empirical determination of instability strip
limits for SX~Phe stars as a function of metallicity. We detected
two additional SX~Phe stars (V62 and V64), which, however, are field
objects. Their observed magnitudes indicate that they are located in
the Galactic halo or the outskirts of the bulge, far behind the cluster.

The variable V52 is a proper motion member of M4. Located ~0.1
mag to the red of the main sequence, it can be called a red
straggler. In Fig. 9 we show its time-domain and phased light
curves. The average luminosity of V52 varies from season to season.
A superimposed periodic variability with $P\approx0.78$~d is also
observed. Both the shape of the light curve and its coherence over
the interval of 14 years indicate that we are dealing with a binary
star. The observed periodic modulation may be due to the ellipsoidal
effect. On UT May 3, 2006 a flare of V52 was observed, lasting for
about 90~min and reaching an amplitude $\Delta V= 0.11$~mag
(Fig.~9). Optical flares are often observed in close binaries with
chromospherically active components. The chromospheric activity
should be accompanied by an X-ray emission. However, none of the
X-ray sources detected in M4 by Bassa et al. (2004) coincides with
V52. Since the variable is located $\sim$1.0 mag above the main
sequence, it cannot be composed of two main sequence stars. As such,
it deserves a spectroscopic follow-up aimed at the determination of
its nature.

The variable red giant V56 is a proper motion member of M4. As 
can be seen in Fig. 10, in addition to seasonal variations of the
average luminosity it exhibits a low amplitude variability on a time
scale of $\sim$10 days. The star may be associated with one from the
close pair of X-ray sources detected by Bassa et al. (2004). The
optical variability of this object may indicate its binary nature.
If V56 is indeed a binary, then the rapid rotation of the giant
component may result in substantial chromospheric activity and
associated X-ray emission. The star was identified as a radial
velocity variable by Sommariva et al. (2009; object No. 34848 in
their Table 4). Two radial-velocity measurements performed 80 days
apart yielded 66.75$\pm$0.12 and 64.31$\pm$0.07 km/s. These values
both confirm the variability of V56 and prove that it is a
radial-velocity member of M4.

The hot subdwarfs V46 and V63 are proper motion members of M4. Both
the stars have blue $U-B$ colors (Mochejska et al. 2002). The binary
nature of V46 was discussed in some detail by O'Toole et al. (2006).
Our data show that the light curve of this variable was very stable
over the time interval of 14 years (Fig. 7). The $V$-band luminosity
of the second subdwarf was systematically increasing from 17.71 mag
in 1995 to 17.68 mag in 2009, however we did not detect any
periodicity in its light curve.

\section{Blue and yellow stragglers}
Our accurate photometry supplemented with proper motion memberships
of Zloczewski et al. (2012) allows the selection of a very clean
sample of blue and yellow stragglers (BSs and YSs) belonging to the
cluster. Fig.~11 shows the relevant region of the CMD, only proper
motion members of M4 are shown. For reference we also plot
a 2.5 Gyr isochrone with ${\rm[Fe/H]}=-1.2$ from the Dartmouth
data base (Dotter et al. 2008). To match the lower main sequence the
isochrone was shifted assuming reddening $E(B-V)=0.392$ mag and
distance modulus $DM=12.81$ mag. This isochrone is a good fit to the 
lower main sequence of M4 on our
CMD. The sequence of BSs emerges from the main sequence below the
cluster turnoff at $V\approx 17.3$ mag and extends to $V=14.6$ mag.
It includes 38 BSs with $B-V<0.81$ and eight YSs located above 
the subgiant branch. As it can be seen in Fig. 11, masses
of these objects range from 0.8 to to 1.4~M$_\odot$, where
0.8~M$_\odot$ is the turnoff mass of M4 (Kaluzny et al. 2013 and references
therein). This estimate is based on the assumption that BSs and YSs follow 
the standard relation between mass, color and luminosity.

The stragglers are thought to originate either as the products of the
evolution of close binary systems or via stellar collisions and
direct merging (McCrea 1964, Hills and Day 1976). The first scenario
may lead to a coalescence of the components, but an end product in the
form of a detached binary emerging from the mass exchange phase is
also possible. 

Coalescence should result in fast rotation of the merger product. 
Lovisi et al. (2010) obtained $Vsini$ values for 20 BSs from M4. 
They found $V_{rot}sini>50$ km/s for eight "fast rotators" and 
lower values of $V_{rot}sini$ for the remaining 12 objects.
Our sample of stragglers includes 16 stars from Lovisi's survey. Seven 
of these are fast rotators and proper motion members of the cluster.
Lovisi's BSs 2000121, 43765 and 2000085 correspond to V53, V61 and V68, 
respectively. For the remaining objects with $V_{rot}sini>50$ km/s we 
found no evidence for photometric variability. Lovisi's BS 2000106 
turned out to be a field star. The remaining eight "slow rotators" from
Lovisi's list are proper motion members of M4. 
Fig. 11 suggests that the fast rotators are relatively unevolved in the 
sense of stellar evolution (they belong to the bluest BSs).  
This in turn may suggest that  they are products of a recent (in case of 
V53 even ongoing) mass exchange in binary systems. This conjecture can 
be verified by time-series spectroscopic observations of our BS/YS sample.

Eclipsing BSs have been found in several clusters, however the membership
status remains to be established for many of them. Those with
confirmed cluster membership include Algols NJL5~=~ V192 in Omega~Cen
(Niss et al. 1978; Bellini et al. 2009), V228 in 47~Tuc (Kaluzny et
al. 2007) and V60 in M55 (Rozyczka et al. 2013), as well as a number
of contact systems. We have examined the population of contact
binaries in seven clusters with proper motion data to find that just
15 of them are blue stragglers, of which 10 belong to the very
massive and atypical cluster $\Omega~Cen$ (Kaluzny and Rozyczka
2013). We conclude that in nearby well studied clusters a
few percent of BSs are eclipsing binaries. M4 seems to follow this
trend: among the 45 BS/YS in M4 there is just one eclipsing binary, the
contact system V53.

The accuracy of our photometry is sufficient for us to  detect periodic variability
at a milli-magnitude level for stars with $V<17$ mag and periods up to
a few days,  as expected in close binaries with an ellipsoidal effect. We
failed to find any such variability. However, this does not exclude the possibility
that the potential binary stragglers are well detached. To verify
such a hypothesis, a radial velocity survey of M4 is needed, similar
to the spectroscopic study of the old open cluster NGC~188 which led
to the detection of 15 binary blue stragglers with orbital periods ranging 
from a few days to $10^{4}$ days (Geller \& Mathieu 2012). 
%The frequency
%of ``hard'' (i.e. tight) binaries among them is as high as $76\pm 19\%$, seeming to
%be significantly higher than observed among main sequence stars of the cluster.\\
%Finally, we note that
Eight binary BSs in NGC~188 are SB1 systems whose secondary
components have masses estimated at 0.6~M$_{\odot}$ (Geller \& Mathieu
2012). It is natural
to assume that such objects are white dwarfs. If this is true, then the 
essential factor responsible for
the formation of those blue stragglers must have been  mass exchange.
One may wonder if binaries hosting white dwarfs are present also
among BS/YS in M4.

\section{Summary}

We monitored M4 photometrically from 1995 till 2009, obtaining over
3000 CCD images of the cluster. Our $V$-band photometry is accurate
to within 5 mmag at the turnoff ($V$ = 17 mag), and to within 50 mmag
at $V$ = 20 mag. The collected data are accurate enough to
measure the differential reddening $\delta E(B-V)$ across the surveyed
field, which varies between $-0.053$ and $+0.011$ mag (with negative
values corresponding to higher reddening). $\delta E(B-V)$ is specified
with respect to a reference subfield, for which we have found $E(B-V)=0.392$
mag using the low-extinction globular cluster NGC~6362 as an intermediary.
Accounting for $\delta E(B-V)$ and using the membership catalog of
Zloczewski et al. (2012), we have derived a very clean color-magnitude
diagram of M4 with sharply defined principal features and a clearly visible
binary main sequence. We also selected a clean sample of a few dozen of 
blue and yellow stragglers. As proper-motion members of M4, they 
can be used for a spectroscopic study which might shed light on the origins
of these populations of stars. 

We have obtained light curves of 22 variables, ten of which are
newly detected. Accurate periods are found for all but two
variables. Nineteen variables are firm or likely proper-motion
members of the cluster. Among these there are four detached binaries,
seven contact binaries, two subdwarfs, one binary which may be
either semidetached or contact, and two SX-Phe pulsator. The
remaining three cluster variables (V52, V56 and V58) are most likely
binaries. The binary nature of V56 is suggested by seasonal
variations of the average luminosity and a low amplitude variability
on a time scale of $\sim$10 days. Such a behavior can be expected
from a red-giant primary, whose rapid synchronous rotation results
in a high chromospheric activity.  In the
case of V52 a periodic modulation of the luminosity with
$P\approx0.78$~d is observed, which may be attributed to the
ellipsoidal effect. The light curve of V58 was obtained by Kaluzny
et al. (2012), who identified this star with the X-ray source CX1
from the list of Bassa et al. (2004). They argue that this system is
composed of a neutron star and a low-mass companion which has lost
most of its hydrogen envelope.

Besides V58, another four cluster variables are either firm or likely optical
counterparts of X-ray sources: V56 seems to be associated with CX5 or CX9 from
the same list, and the optical counterparts of CX13, CX15 and CX30 are, respectively,
eclipsing binaries V48, V49 and V65. V52 is not associated with any known X-ray source,
however on May 3rd, 2006 it flared for about 90~min at an amplitude $\Delta V= 0.11$~mag.
This type of activity also should be accompanied by an X-ray emission. Among the
three field objects which complete the sample of detected variables there are
are two SX-Phe pulsators and a contact binary.

We find that all but one of the contact (EW) systems show
season-to-season variations of the average luminosity, with an
amplitude of more than 0.1~mag. Such  behavior
can be explained by magnetic cycles five to ten years long.
Interestingly, their periods are similar to that of the
solar cycle despite a huge difference in rotation rates. Another
interesting finding is that the contact binaries appear exclusively
among evolved stars, which suggests that at least in globular
clusters the principal factor enabling EW systems to form from close
but detached binaries is stellar evolution. The estimated
frequency of occurrence of EW systems in M4, 0.12$\pm$0.05\%, is
consistent with the value of 0.2\% observed in the solar neighborhood. Two
contact binaries with total eclipses, V48 and V53, have very low
mass ratios of $q=0.148$ and $q=0.060$, respectively. The latter
value poses a challenge for the current theory, according to which
$q=0.09-0.08$ is the lower limit for tidally stable contact
binaries.

The detached binaries V65 and V66, located at the turnoff, were additionally
subject to a thorough spectroscopic survey. Their detailed analysis including
the determination of absolute parameters and ages can be found in Kaluzny et al.
(2013). 
The light curves of all variables as well as other data presented in this
paper are available online at http://case.camk.edu.pl/results/index.html.

\Acknow{
We thank Loredana Lovisi for sending us the coordinates of M4 stragglers
with measured rotation rates.
JK and MR were partly supported by the grant NCN 2012/05/B/ST9/03931
from the Polish Ministry of Science. IBT was supported by NSF grant
AST-0507325.}

\newpage

% Fig 1
\begin{figure}[htb]
\centerline{\includegraphics[width=120mm, bb= 18 221 494 497,clip]{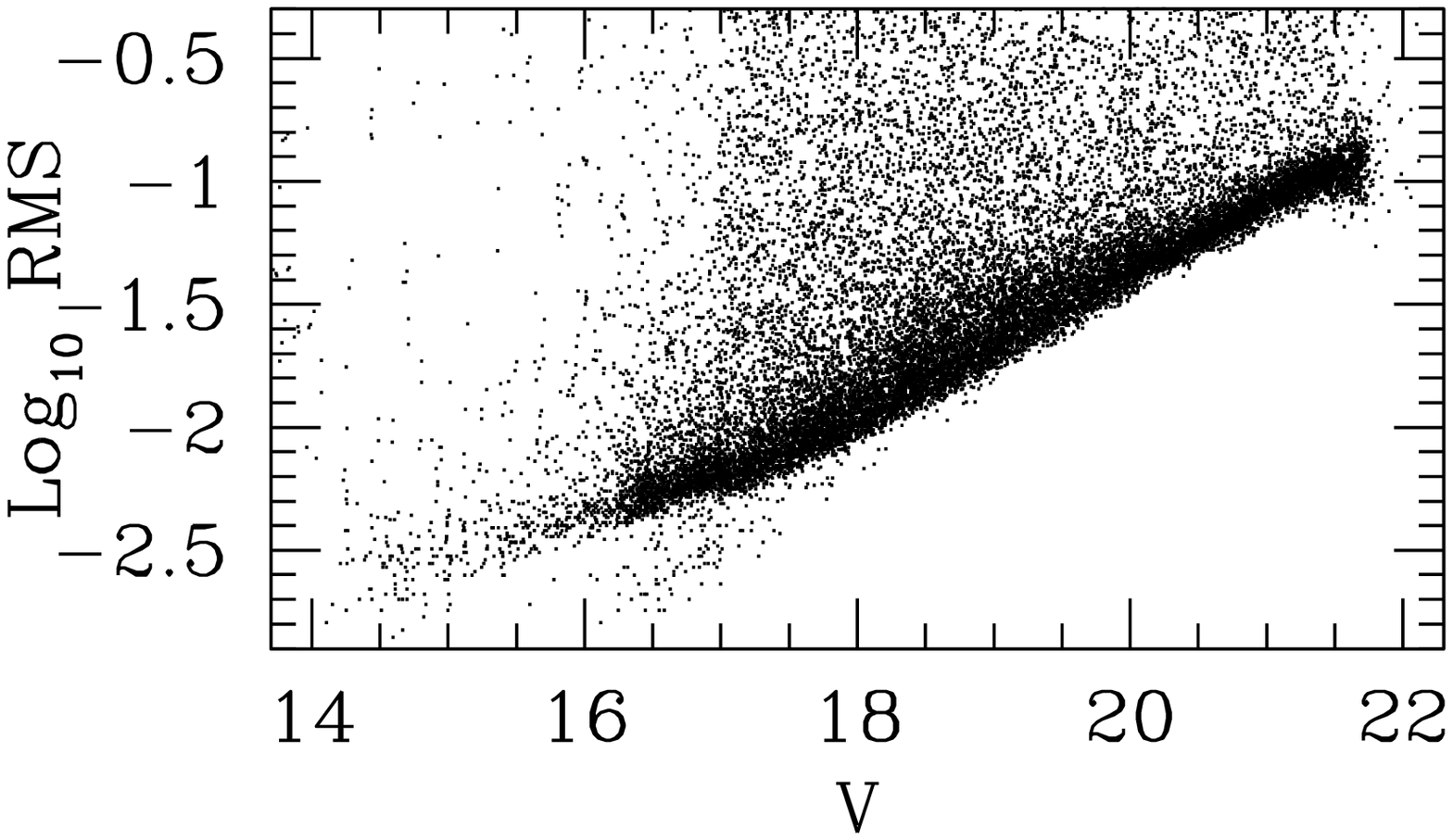}}
\caption{\small
Photometric measurements of stars in the M4 field. Standard
deviation is plotted vs. average $V$-magnitude.}
\end{figure}

% Fig 2
\begin{figure}[htb]
\centerline{\includegraphics[width=120mm, bb= 82 199 495 462, clip]{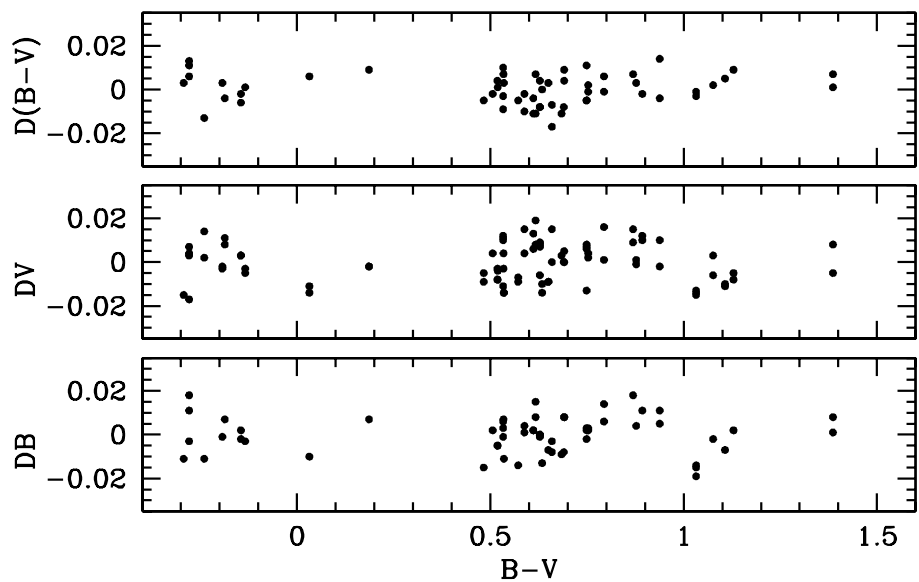}}
\caption{\small
Residuals resulting from the transformation defined by Equations
(1)-(3). Difference between original and recovered magnitudes of
Landolt standard stars are plotted as a function of color. }
\end{figure}

% Fig 3
\begin{figure}[htb]
\centerline{\includegraphics[width=120mm, bb= 39 193 442 567,clip]{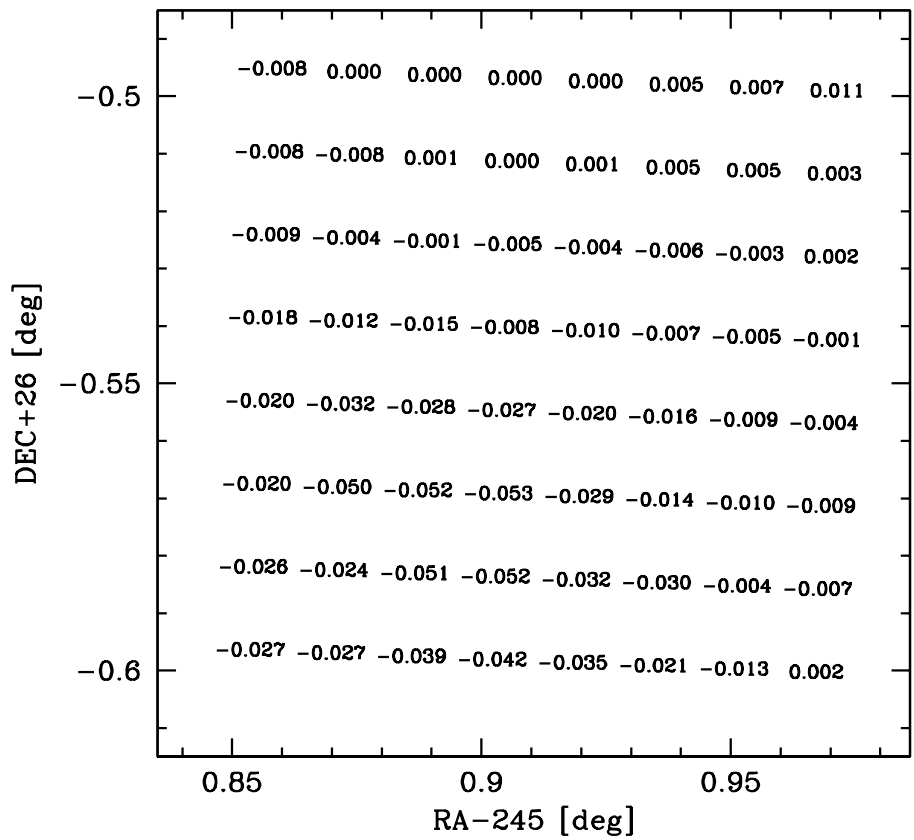}}
\caption{\small
Differential extinction $\delta E(B-V)$ in the M4 field as a
function of J2000 equatorial coordinates.}
\end{figure}

% Fig 4
\begin{figure}[htb]
\centerline{\includegraphics[width=120mm]{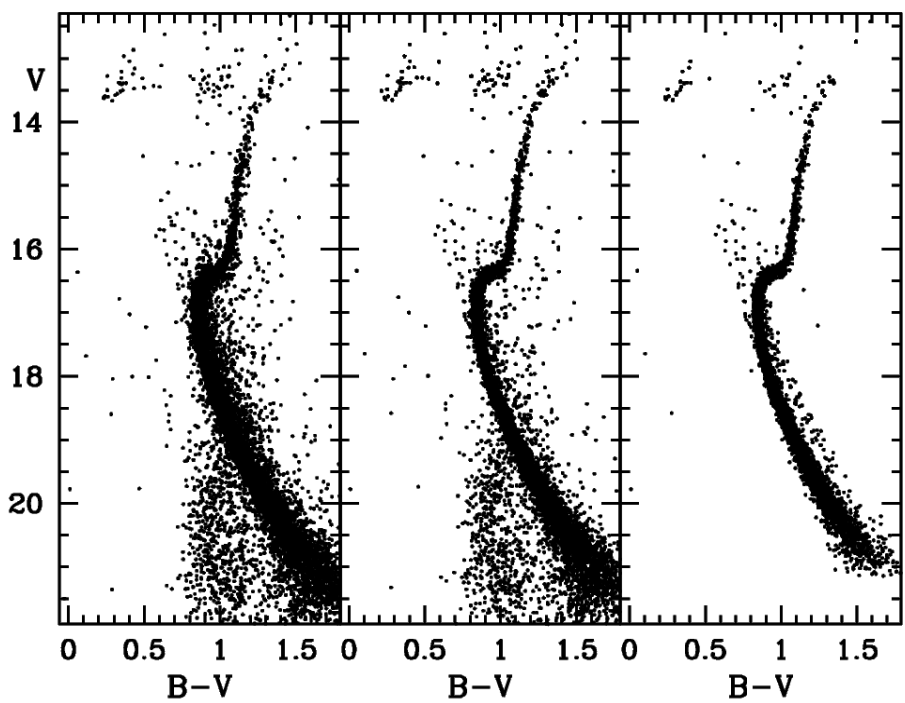}} \caption{\small
CMD of M4. In the case of variable objectes magnitudes and colors 
correspond to the phases at which the reference frames were taken. 
Left: all stars, no correction for differential
reddening. Middle: stars with large errors in photometry are
removed; remaining stars are corrected for differential reddening.
Right: same as middle for proper motion members of the cluster
(membership was determined for $V<21$ mag only).}
\end{figure}

% Fig 5
\begin{figure}[htb]
\centerline{\includegraphics[width=120mm, bb= 54 340 578 450, clip]{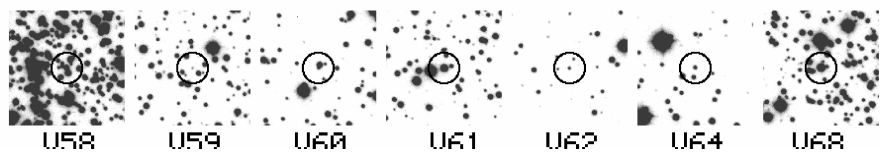}}
\caption{\small
Finding charts for variables V58-61, V62, V64 and V68. Each chart is 30
arcsec on a side, with north up and east to the left. See text for references 
to charts published elsewhere. }
\end{figure}

%Fig 6
\begin{figure}[htb]
\centerline{\includegraphics[width=120mm, bb= 43 30 565 742, clip]{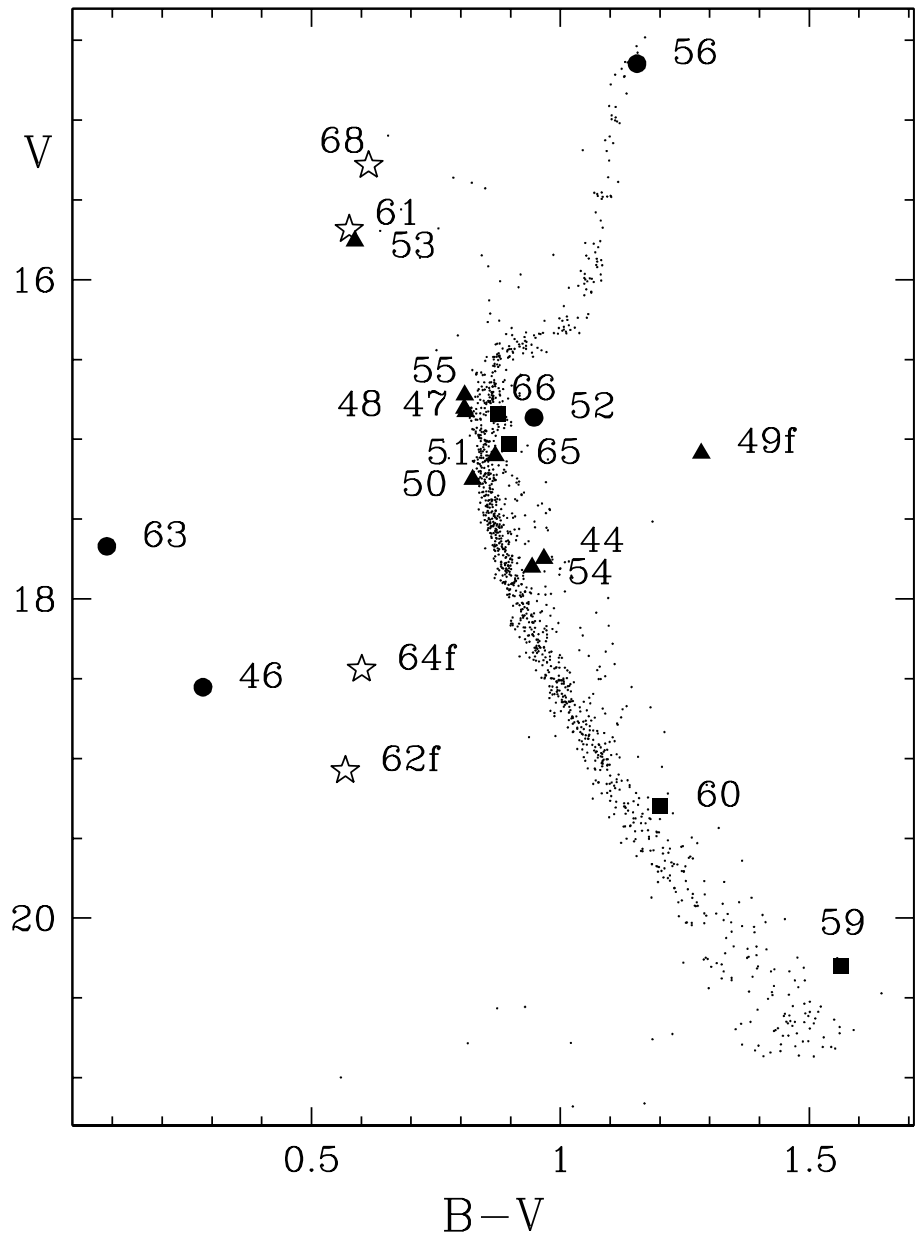}}
\caption{\small
CMD of M4 with positions of variables detected within the CASE program.
Magnitudes and colors of the variables are taken from Table 2, i.e.  
they either correspond to the phase of the maximum brightness or are median 
values from all measurements.
Squares: detached binaries; triangles: contact binaries; stars: SX Phe
variables (field objects are marked with ``f'').
Background dots: non-variable stars selected from a small section of the
observed field to avoid crowding.}
\end{figure}

% Fig 7
\begin{figure}[htb]
\centerline{\includegraphics[width=120mm, bb= 53 149 521 707, clip]{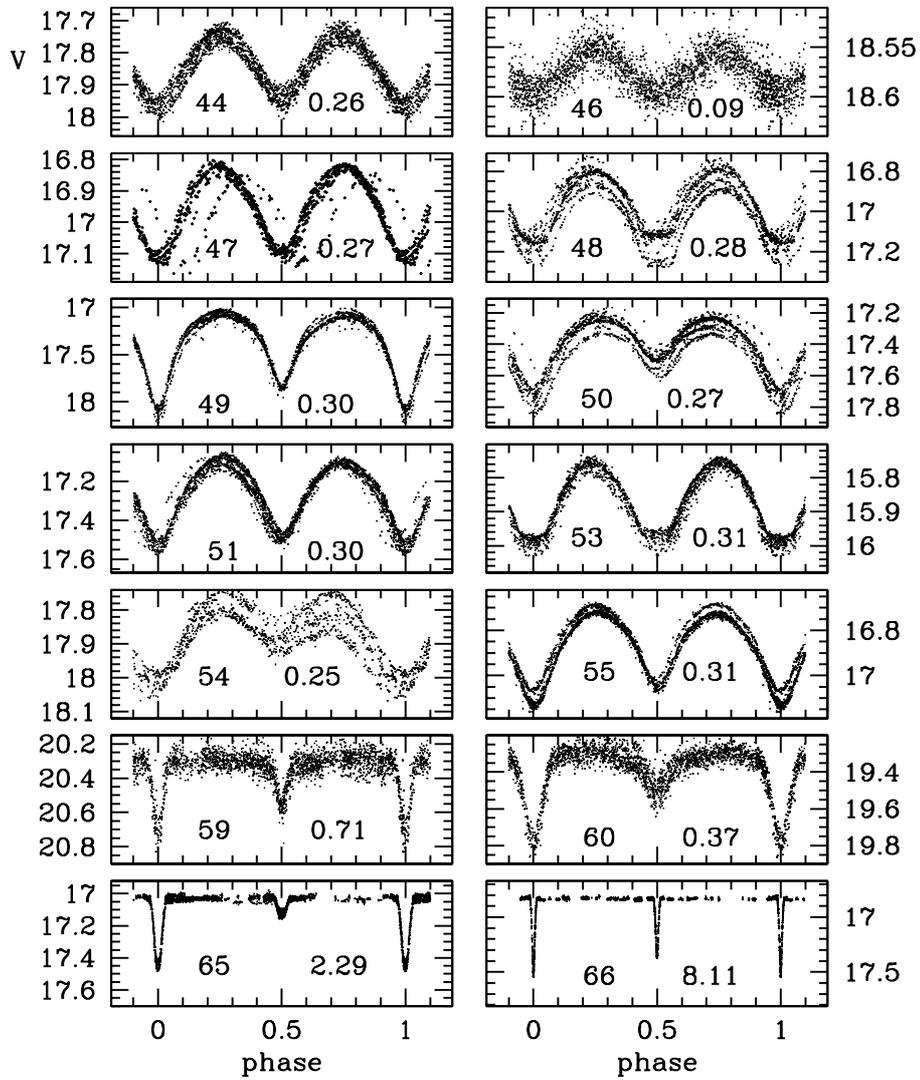}}
\caption{\small
Phased $V$-light curves of eclipsing binaries and the ellipsoidal
variable V46. In each panel the left label gives the name of the variable
(with V omitted to save space), and the right one the orbital period in days. }
\end{figure}

% Fig 8
\begin{figure}[htb]
\centerline{\includegraphics[width=120mm, bb= 52 437 514 707,clip]{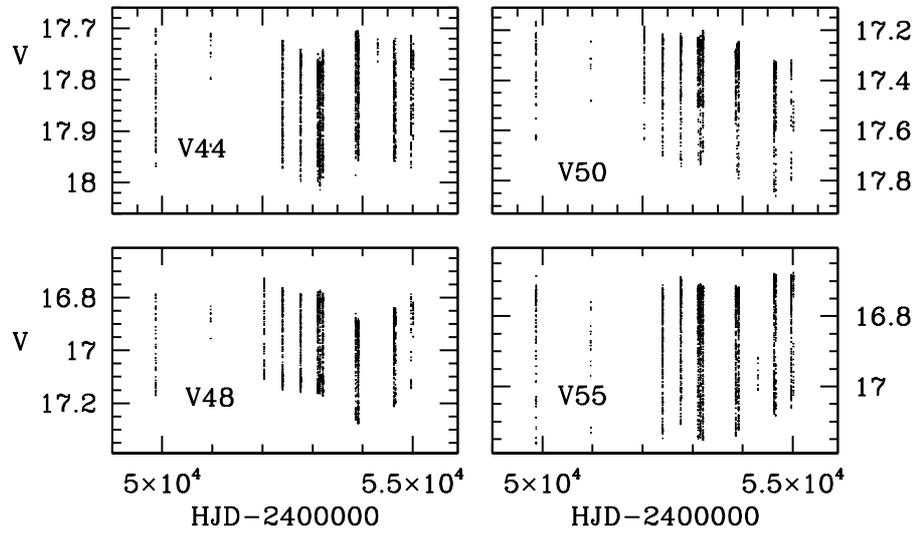}}
\caption{\small
Light curves of stars V44, V48, V50 and V55. }
\end{figure}

% Fig 9
\begin{figure}[htb]
\centerline{\includegraphics[width=120mm, bb= 52 487 530 707, clip]{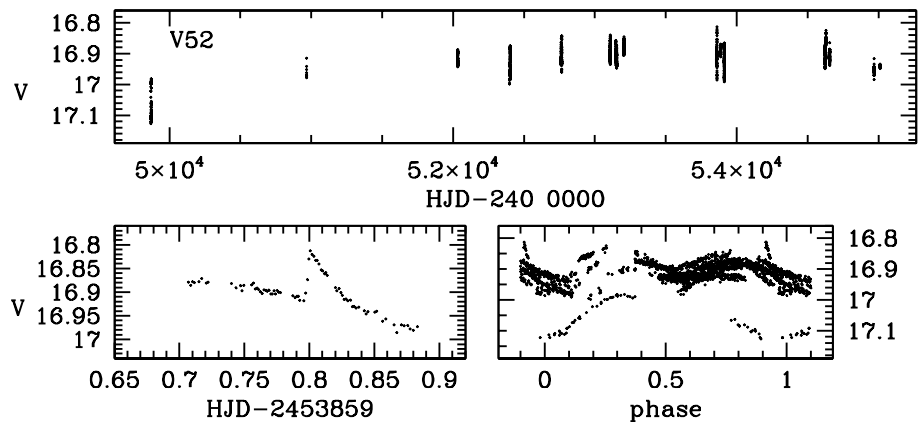}}
\caption{\small
Time-domain and phased light curves of the variable V52 in $V$-band.}
\end{figure}

% Fig 10
\begin{figure}[htb]
\centerline{\includegraphics[width=110mm, bb= 45 486 530 707, clip]{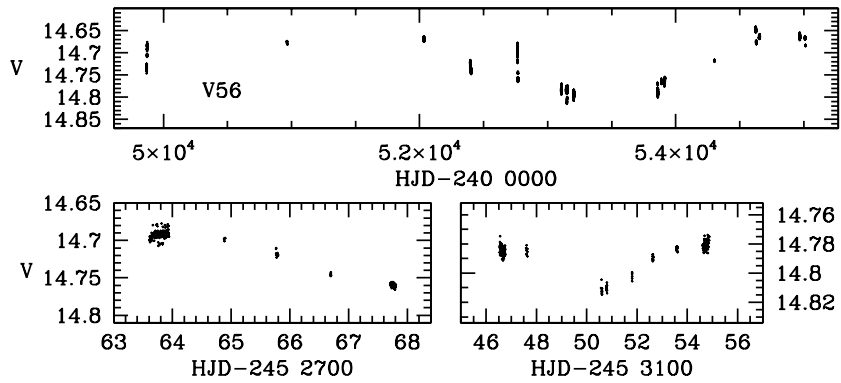}}
\caption{\small Light curves of the variable V56. }
\end{figure}

% Fig 11
\begin{figure}[htb]
\centerline{\includegraphics[width=110mm, bb= 64 175 362 462]{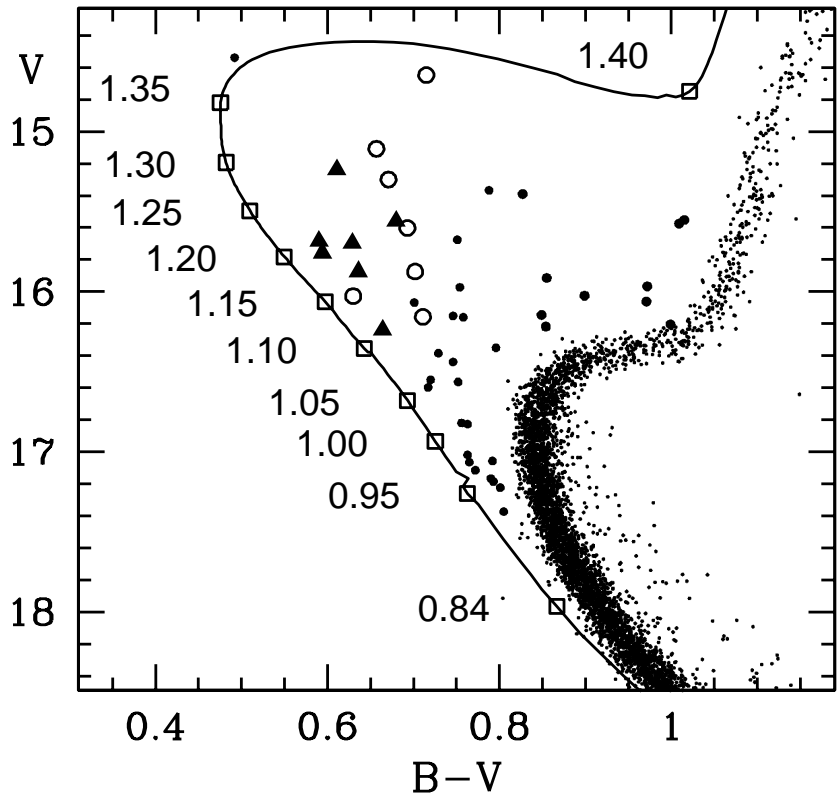}}
\caption{\small Color-magnitude diagram for proper motion members of M4.
Positions of blue and yellow straglers are marked with triangles (fast rotators), 
circles (slow rotators) and large dots (objects with unknown rotation rates).
A 2.5~Gyr isochrone is also shown, with stellar masses marked at selected
locations.}
\end{figure}

\end{document}